\documentclass[twocolumn, aps, prb, superscriptaddress]{revtex4}

\usepackage{amssymb,amsmath}
\usepackage{graphicx, epsfig}

\def \hf{\tfrac{1}{2}}    

\def \ord{\mathcal{O}}

\def\lba{\left(}    \def\rba{\right)}
\def\lbc{\left[}    \def\rbc{\right]}

\newcommand{\bra}[1]{\langle\left.{#1}\right|}
\newcommand{\ket}[1]{\left|{#1}\right.\rangle}
\newcommand{\xpct}[1]{\langle{#1}\rangle}    

\begin{document}

\title{Finite-rate quenches of site bias in the Bose-Hubbard dimer}

\author{T. Venumadhav}
\affiliation{Max-Planck Institute for the Physics of Complex Systems, N\"othnitzer Str.~38, 01187 Dresden, Germany}
\affiliation{Department of Physics, Indian Institute of Technology, Kanpur 208016, India}

\author{Masudul Haque}
\affiliation{Max-Planck Institute for the Physics of Complex Systems, N\"othnitzer Str.~38, 01187 Dresden, Germany}

\author{R. Moessner}
\affiliation{Max-Planck Institute for the Physics of Complex Systems, N\"othnitzer Str.~38, 01187 Dresden, Germany}

\begin{abstract}

For a Bose-Hubbard dimer, we study quenches of the site energy imbalance,
taking a highly asymmetric Hamiltonian to a fully symmetric one.  The ramp is
carried out over a finite time that interpolates between the instantaneous and
adiabatic limits.  We provide results for the excess energy of the final state
compared to the ground state energy of the final Hamiltonian, as a function of
the quench rate.  
This excess energy serves as the analog of the defect density that is
considered in the Kibble-Zurek picture of ramps across phase transitions.
We also examine the fate of quantum `self-trapping' when the ramp is not
instantaneous.

\end{abstract}




\maketitle


\section{Introduction} 
Explicit time evolution of quantum many-particle systems out of equilibrium
has generated intense interest due to unprecedented possibilities for
experimentally accessible non-equilibrium situations, opened up by
developments in laser-cooled atomic clouds and in mesoscopic systems.  One
theme has been the response to an instantaneous `quench', where a physical
parameter is suddenly changed to a different value.
%
While instantaneous quenches are more convenient to analyze, a change of
parameter can of course be performed at any rate.  At the other extreme from
the instantaneous quench, one can make the parameter sweep
\emph{adiabatically}, in which case the system reaches the ground state of the
final Hamiltonian.
%
%
%
%
%
The Kibble-Zurek theory describes finite-rate ramps,
neither instantaneous nor adiabatic, across phase transitions. \cite{KibbleZurek} 
While the original interest concerned thermal phase transitions, there has
been a recent surge of interest in finite-rate traversals of \emph{quantum}
phase transitions in lattice Hamiltonians
(Refs.\ \onlinecite{finiteRateQuenches_othersystems,
finiteRateQuenches_othersystems_residualenergy,
CucchiettiDamskiDziarmagaZurek_BoseHubbardDynamics_PRA07} and references
therein).  One important context is the Bose-Hubbard Hamiltonian, where
finite-rate ramps of the Hubbard interaction $U$ have been analyzed,
\cite{Polkovnikov_PRA03, SchuetzholdUhlmannXuFischer_PRL06,
CucchiettiDamskiDziarmagaZurek_BoseHubbardDynamics_PRA07} motivated by an
influential experiment quenching across the Mott-superfluid transition. 
\cite{GreinerBloch_Nature2002}

In this work, we analyze finite-rate ramps in a Bose-Hubbard dimer.  For
far-from-equilibrium issues where few standard theoretical techniques exist,
it is of obvious interest to look at finite clusters because of near-exact
solvability.  One can hope to understand the non-equilibrium dynamics in some
detail.  Such detailed results for clusters clearly provide an invaluable
background for the emerging field of non-equilibrium dynamics in macroscopic
(many-site) quantum systems.
In addition, the bosonic dimer is an important model system by itself, and its
physics is relevant in diverse contexts.  
In recent years several experiments have achieved two-site bosonic systems in
cold-atom setups. \cite{AlbiezOberthaler_PRL05, ShinKetterlePritchard_PRL05,
GatiOberthaler_JPhysB07, FoellingBloch_Nature07, CheinetBloch_PRL08}  The
quenches we study could be implemented in such a setup, \emph{e.g.}, through
slower ramps in the experiment of Ref.~\onlinecite{AlbiezOberthaler_PRL05}.
The Bose-Hubbard dimer can be mapped onto a single-spin Hamiltonian, 
\cite{MilburnCorneyWrightWalls_PRA97,
RaghavanSmerziKenkre_PRA99,RaghavanSmerziFantoniShenoy_PRA99,
KalosakasBishopKenkre_JPhysB03,KalosakasBishopKenkre_PRA03,
Salgueiro-etal_EPJD07, TonelLinksFoerster_JPA05}
very similar to that governing single-molecule magnetic experiments, where
finite-rate ramps have received some attention.
\cite{WernsdorferSessoli_Science99,
Wernsdorfer-etal_LZ-in-molecularmagnet_EPL00}
Finite-rate quenches 
have also been reported with a Josephson junction arrangement, 
\cite{Johansson-etal_LandauZener_fluxqubit_PRB09} which is closely related to
a bosonic dimer.

For $N$ bosons in two sites, the Hamiltonian is
\begin{multline*}
H ~=~ ~-\frac{K}{2} \left( a_1^{\dagger}a_2 ~+~ a_2^{\dagger}a_1\right)  
~+~ \frac{U}{2} \sum_{i=1}^{2} n_i(n_i-1)  
\\ ~+~  \frac{\delta}{2} (n_1-n_2)
\end{multline*}
with $n_i=a_i^{\dagger}a_i$ and $\xpct{n_1}+\xpct{n_2}=N$.  
%

Quenches of the interaction $U$ in the dimer have been considered in
Refs.\ \onlinecite{CucchiettiDamskiDziarmagaZurek_BoseHubbardDynamics_PRA07,
Polkovnikov_PRA03, SchuetzholdUhlmannXuFischer_PRL06,
interactionQuench_instantaneous}.
%
%
However, for the dimer, quenches of the site bias parameter $\delta$ are also
of natural interest.  Instantaneous quenches of $\delta$ from an imbalanced to
a symmetric situation, \emph{i.e.}, $\delta(t) = \delta_0\,\theta(-t)$, have
been studied as a quantum realization of \emph{self-trapping}.  
\cite{MilburnCorneyWrightWalls_PRA97,
RaghavanSmerziKenkre_PRA99,RaghavanSmerziFantoniShenoy_PRA99,
KalosakasBishopKenkre_JPhysB03,KalosakasBishopKenkre_PRA03,
Salgueiro-etal_EPJD07, TonelLinksFoerster_JPA05}
%
Classical self-trapping, described by the two-mode Gross-Pitaevskii equation,
is the persistence of imbalance in the state despite the final Hamiltonian
being unbiased.  \cite{AlbiezOberthaler_PRL05, ClassicalSelfTrapping}
%
%
In the quantum case, the relative number imbalance $z=\xpct{n_1-n_2}/N$
oscillates after the instantaneous quench, with decay and long-time revivals
of the oscillations.  For small $\delta_0$, the imbalance oscillations are
around zero.
%
However, for large enough values of $\delta_0$ and $NU/K$, after a fast quench
the mean value of the oscillating $z(t)$ does not relax to zero within any
reasonable time scale.  \cite{MilburnCorneyWrightWalls_PRA97,
RaghavanSmerziKenkre_PRA99,RaghavanSmerziFantoniShenoy_PRA99,
KalosakasBishopKenkre_JPhysB03,KalosakasBishopKenkre_PRA03,
Salgueiro-etal_EPJD07, TonelLinksFoerster_JPA05}
%
The astronomically long relaxation times (``quantum self-trapping'') can be
understood in terms of extremely small energy splittings in the eigenvalue
spectrum of the final Hamiltonian.  For the instantaneous quench, the various
relationships between timescales of dynamical features and energy scales of
the $\delta=0$ spectrum has been discussed in some detail in the literature.  
\cite{MilburnCorneyWrightWalls_PRA97,
RaghavanSmerziKenkre_PRA99,RaghavanSmerziFantoniShenoy_PRA99,
KalosakasBishopKenkre_JPhysB03,KalosakasBishopKenkre_PRA03,
Salgueiro-etal_EPJD07, TonelLinksFoerster_JPA05}


\begin{figure}
\centering
 \includegraphics*[width=0.95\columnwidth]{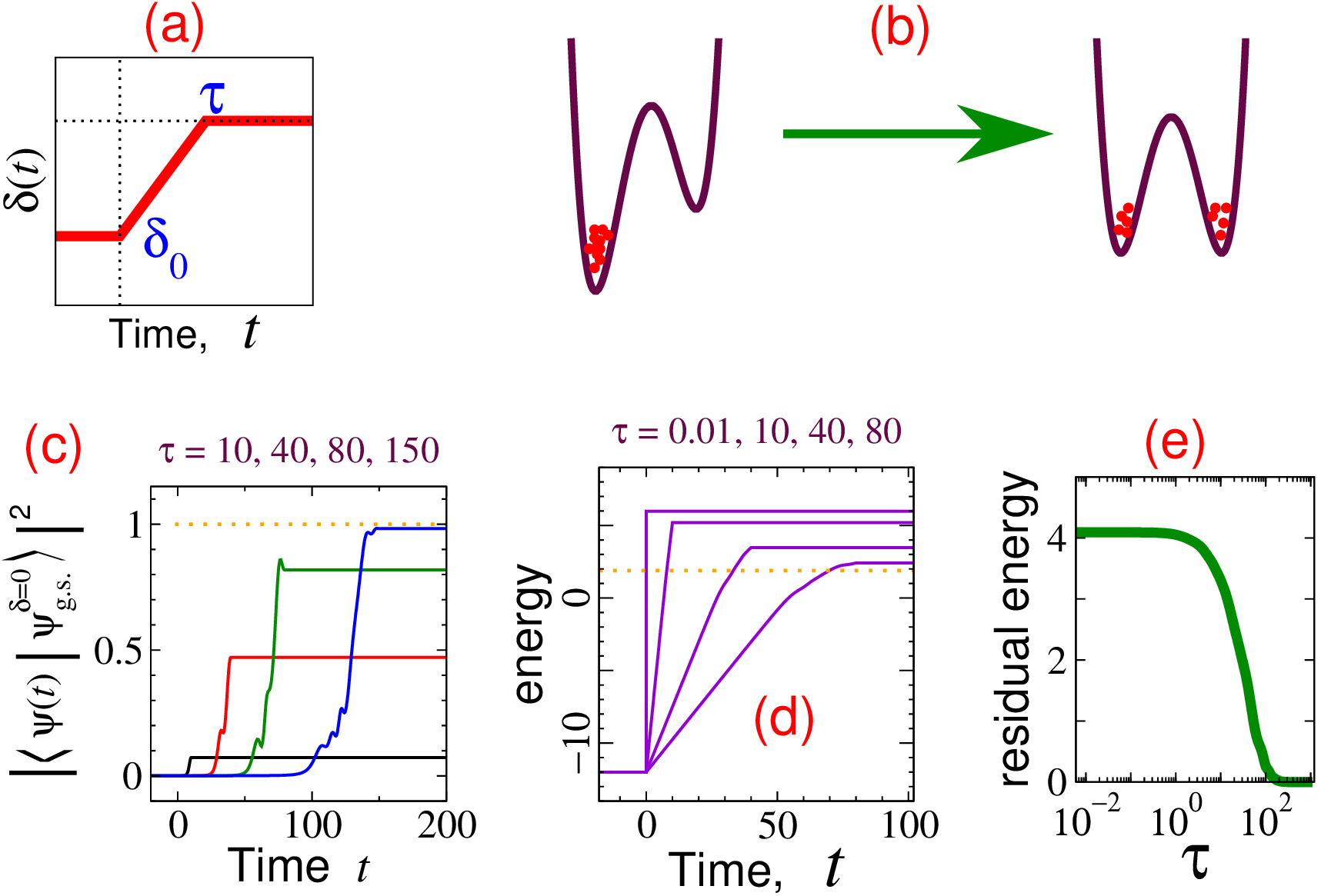}
\caption{ \label{fig_energyEvolution}
(Color online.)  (a) and (b) Form of quench analyzed in this work.  
Lower panels illustrate the approach to adiabaticity with increasing $\tau$,
for ($N$,$K$,$\delta_0$)=(4,\,0.2,\,9).  
(c) overlap with ground state of final Hamiltonian.  Curves from lowest to
highest final values correspond to $\tau=$ 10, 40, 80, 150.
(d) Evolution of energy:  curves from top to bottom are $\tau=$ 0.01, 10,
  40, 80.  Dotted horizontal line is the ground state energy of the final
  Hamiltonian.  The final excess energy over this dotted line is the residual
  energy, which decreases with
  $\tau$ (e).
}
\end{figure}

We use $U=1$, measuring energy [time] in units of $U$ [$\hbar/U$].  We focus
mostly on the regime $K<<U$.
We will analyze finite-rate quenches of $\delta$, of the form
\[
\delta(t) = -\delta_0 \; \theta(-t) ~+~ (-\delta_0/\tau) ( \tau - t)\;\; \theta(t)\;\;
\theta( \tau-t) \; ,
\]
as illustrated in Figs.~\ref{fig_energyEvolution}a, \ref{fig_energyEvolution}b.
We consider the whole range from $\tau=0$ (instantaneous) to
$\tau\rightarrow\infty$ (adiabatic).  
The initial ($t=0$) state is taken to be the ground state of the initial
Hamiltonian with $\delta=-\delta_0$.
The initial asymmetry $\delta_0$ is
taken to be large enough that the bosons are initially concentrated almost
entirely on  site 1, \emph{i.e.}, the wavefunction is dominated by
$\ket{N,0}$.  The final ground state is dominated by $\ket{N/2,N/2}$ but the
system does not reach this ground state unless the quench is truly adiabatic.
Fig.~\ref{fig_energyEvolution}c demonstrates how larger-$\tau$ quenches are more
nearly adiabatic, through the temporal evolution of overlaps with the
$\delta=0$ ground state.

In Kibble-Zurek theory, one uses the \emph{defect density} in the final
ordered state to quantify the deviation from adiabaticity.  In few-site
clusters, defects are not easily defined, nor are phase transitions or
ordering.  However, there is a natural quantity that serves an analogous role,
namely, the final energy after the quench, $E_{t>\tau} =
\bra{\psi_{t>\tau}}H^{\delta=0}\ket{\psi_{t>\tau}}$.  In an adiabatic sweep,
the final energy is the ground state energy $E^{\delta=0}_{g.s.}$ of the final
Hamiltonian.  The excess energy over $E^{\delta=0}_{g.s.}$ (\emph{residual
energy}) measures the deviation from adiabaticity.  
\cite{finiteRateQuenches_othersystems_residualenergy}  The interpolation
between instantaneous and adiabatic limits is illustrated through energy
evolution in Fig.~\ref{fig_energyEvolution}d, and through residual energies in
Fig.~\ref{fig_energyEvolution}e.

Our main results concern the dependence of the residual energy, $\Delta{E} =
E_{t>\tau} - E^{\delta=0}_{g.s.}$, on the quench time $\tau$.  
For near-instantaneous quenches (small $\tau$), the energy deviation from the
instantaneous limit is found to scale as $\tau^2$.
We analyze larger-$\tau$ quenches through a multi-crossing Landau-Zener
\cite{LandauZener} scenario, and derive concise expressions for $\Delta{E}$ in
an intermediate-$\tau$ regime as well as in the near-adiabatic (very large
$\tau$) limit.

While we focus on $\Delta{E}$, our analysis can in principle be adapted to
treat other observables.
Since instantaneous $\delta$-quenches are associated with the self-trapping
phenomenon, we also ask whether and how self-trapping survives when the ramp
rate is finite.

\section{Slow quenches} \label{sec_largetau}

\begin{figure}
\centering
 \includegraphics*[width=0.95\columnwidth]{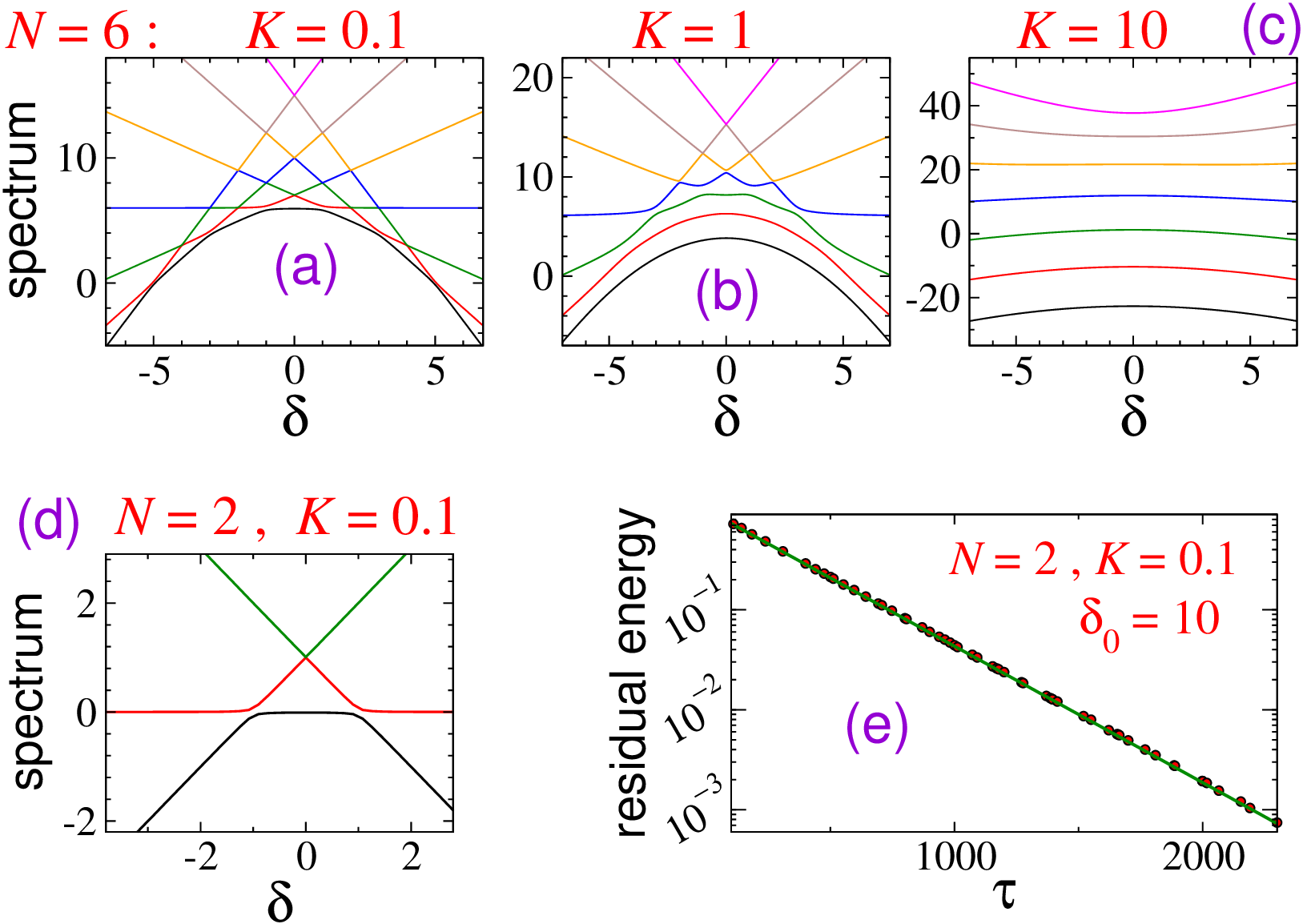}
\caption{ \label{fig_crossings_excessenergy}
(Color online.)  (a-c) energy spectra for $N=6$ bosons.  For small $K$ and
 even $N$, the lowest state goes through $N/2$ avoided level crossings on
 either side of $\delta=0$, at $\delta = \pm(2x-1)U$, with $x=1,2,...,N/2$.
 (d) $N=2$ bosons. (e) Residual energy; dots are exact values and line is
 theory.
}
\end{figure}

Our analysis for large $\tau$ relies on the avoided level crossing structure
of the problem.  Fig.~\ref{fig_crossings_excessenergy}a-c shows that the level
crossing structure is only relevant for small $K$.  Our analytic treatment for
slow quenches is focused on this parameter region, $K\ll{U}$ (
Fig.~\ref{fig_crossings_excessenergy}a).  Therefore, we can regard the quench
problem as one of traversing a sequence of well-separated avoided level
crossings.  This scenario would clearly not be applicable for situations such
as those in Figs.\ \ref{fig_crossings_excessenergy}b,
\ref{fig_crossings_excessenergy}c.

We start with the simplest case of two bosons.  There is only one level
crossing encountered during the quench, at $\delta\approx-1$, where the
$\ket{2,0}$ and $\ket{1,1}$ states are mixed
(Fig.~\ref{fig_crossings_excessenergy}d).  The Hamiltonian within this space
is
$H = \begin{pmatrix}
1+\delta(t) & -K/\sqrt{2} \\  -K/\sqrt{2} & 0 
    \end{pmatrix}$. 
The Landau-Zener formula \cite{LandauZener} gives the probability of
excitation at this level crossing to be $p=e^{-2\pi\gamma}$, where
$\gamma=(K/\sqrt{2})^2/\dot{\delta}$.
At the end of the quench, the energy has a contribution of weight $(1-p)$ from
the ground state (energy $\approx0$) and a contribution of weight $p$ from the
two higher levels (energy $\approx{U}$). Thus the residual energy is
$\Delta{E}\approx Ue^{-(\pi{K}^2/\delta_0)\tau}$.  This reproduces numerical
calculations for large $\tau$ (Fig.~\ref{fig_crossings_excessenergy}e).

\begin{figure}
\centering
 \includegraphics*[width=0.95\columnwidth]{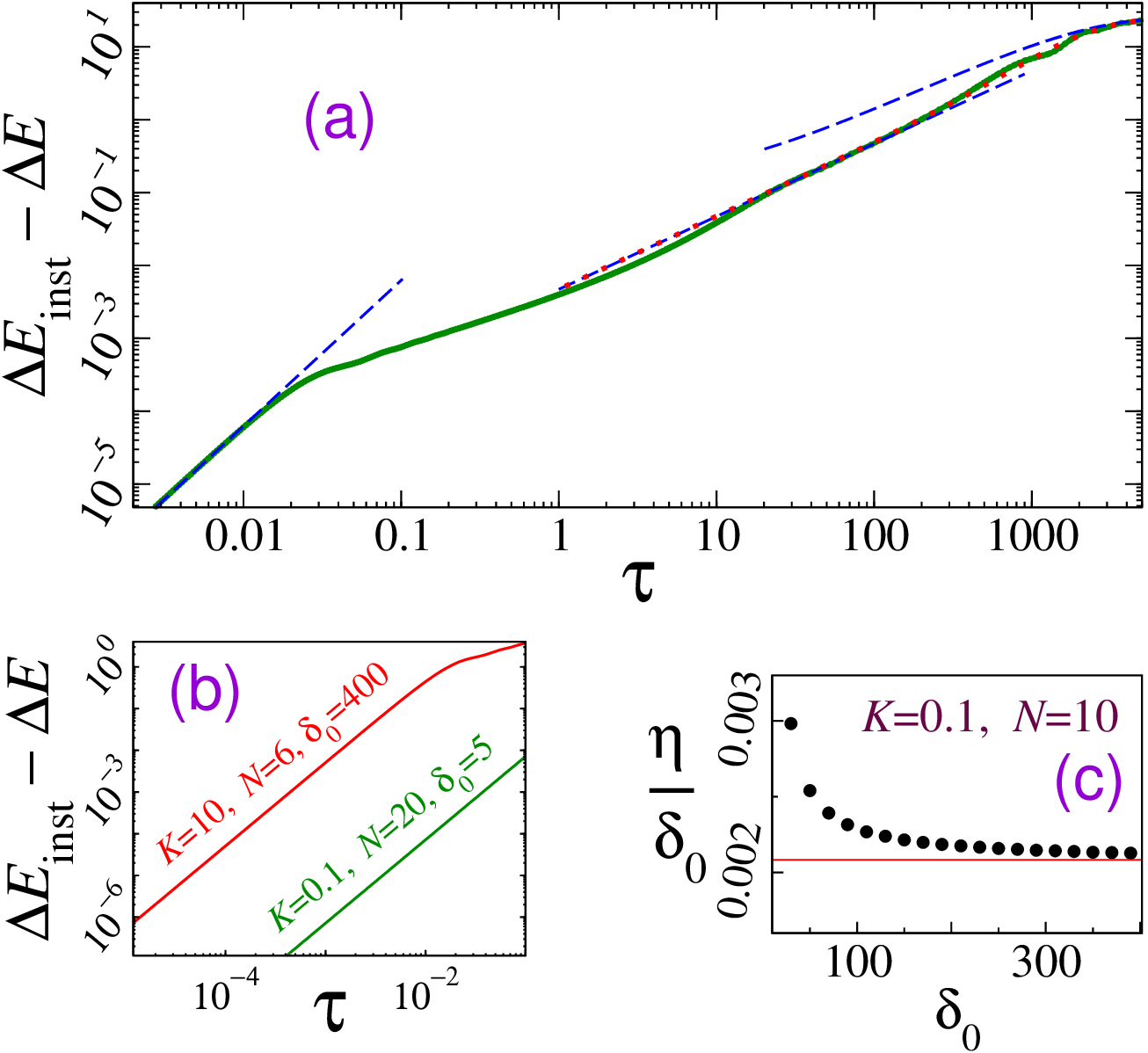}
\caption{ \label{fig_AlltauSmalltau}
(Color online.)  (a) ($N$,$K$,$\delta_0$) = (10,0.1,300).  Thick line: exact
  numerical values.  The three dashed lines are the analytical results for
  small, intermediate and large $\tau$ (Eqs.\ \ref{eq_smalltau2},
  \ref{eq_intermediate1}, \ref{eq_largetau2}).  Eq.\ \ref{eq_largetau1} is
  dotted line interpolating between \ref{eq_intermediate1},
  \ref{eq_largetau2}, but is barely visible because the exact curve almost
  coincides. (The exact curve has some additional oscillations.)
(b) Quadratic behavior, shown here for cases we do not treat analytically,
  large $K$ (upper) and $\delta_0<NU$  (lower).
(c) Coefficient $\eta$ follows Eq.\ \ref{eq_smalltau2} (horizontal line) when
  $\delta_0{\gg}NU$.
%
%
}
\end{figure}

For larger numbers of bosons, the quantum state can take various paths to
$\delta=0$, in a multi-crossing situation such as that shown in
Fig.~\ref{fig_crossings_excessenergy}a.  Fortunately, at small $K$ the higher
crossings are simple to treat because the energy splitting at these points are
of higher than linear order in $K$.  Thus for $K{\ll}U$ one can regard these
as real crossings rather than avoided crossings, so that the excitation
probabilities are unity.  One therefore has to consider only the excitation
probabilities at the $N/2$ crossings involving the lowest energy state.  The
excitation probabilities are 
\[
p_i=\exp[-\tfrac{\pi}{2}i(N-i+1)K^2\tau/\delta_0]
\]
at the $i$-th crossing encountered during the quench; here $i(N-i+1)$ is the
bosonic factor relevant for the coupling between states $\ket{N-i+1,i-1}$ and
$\ket{N-i,i}$.  Any weight going into the upper level at the
($\tfrac{N}{2}-\alpha+1$)'th crossing goes straight on to the final energy
$E_\alpha$, since we neglect further deflection at the higher crossings.  The
final ($\delta=0$) energies are $E_{\alpha} \approx
\tfrac{N(N-2)}{4}+\alpha^2$, with $\alpha = 0,1,,...N/2$.  From this picture,
the final energy is found to be
\begin{multline}   \label{eq_largetau1}
p_1 E_{N/2} + (1-p_1)p_2 E_{N/2-1} + (1-p_1)(1-p_2)p_3  E_{N/2-2} + ....
\\
+ \lbc\prod_{i=1}^{N/2-1}(1-p_i)\rbc p_{N/2} E_1 + \lbc\prod_{i=1}^{N/2}(1-p_i)\rbc  E_0 
\; .
\end{multline}
For moderate values of $N$ for which exact numerical evolution is feasible up
to large times, we find this relationship to work very well at small $K$
(\emph{e.g.}, Fig.~\ref{fig_AlltauSmalltau}a).

There are two situations, corresponding to distinct physical pictures of the
excitation process, where Eq. \eqref{eq_largetau1} reduces to compact forms.
First, when $\tau$ is large enough that one can use the Landau-Zener formula,
but small enough that the $p_i=e^{-a_i\tau}$ are close to unity, the excitation
at the first two crossings deplete the weight, since $p_{1,2}\approx{1}$.
Thus only the top two final levels ($E_{N/2}$ and $E_{N/2-1}$) contribute to
the final energy.  Expanding $p_1=e^{-a_1\tau}$ to linear order, one gets for
this intermediate regime 
\begin{equation}  \label{eq_intermediate1}
 \Delta{E}_{\rm inst.} -  \Delta{E} ~\approx~ (N-1)
 \frac{\pi{K^2}N}{2\delta_0} \tau  \, . 
\end{equation}
Here $\Delta{E}_{\rm inst.}$ is the residual energy for the instantaneous case
($\tau=0$), namely $\bra{\psi_{t=0}}H^{\delta=0}\ket{\psi_{t=0}} -
E^{\delta=0}_{g.s.}$.  We have used 
$\Delta{E}_{\rm inst.} \approx E_{N/2} - E_{0} \approx N^2/4$. (In an
instantaneous quench, the weight would all go to the highest final level.)
One could also extend the above analysis by including the third crossing and
expanding up to $\tau^2$.  Obviously, including further levels and expanding
the exponentials up to higher orders in $\tau$, one eventually gets back the
full expression \eqref{eq_largetau1}.  Note that, although we have obtained a
linear intermediate behavior in $\tau$, this is a nonperturpative result since
it is is based on Landau-Zener probabilities.

Second, at very large $\tau$ the $p_i$ are small, so that
$p_1{\ll}p_2{\ll}...{\ll}p_{N/2}$, because of the bosonic factors $i(N-i+1)$.
Neglecting $p_{i>1}$, one obtains
\begin{equation}  \label{eq_largetau2}
\Delta{E} ~\approx~  p_1{}E_{N/2} - p_1{}E_0  ~\approx~ \tfrac{1}{4} N^2
\exp\lbc-\frac{\pi{N}K^2}{2\delta_0}\tau\rbc \, .  
\end{equation}
Only the lowest and highest energy levels contribute in this case.

Fig.~\ref{fig_AlltauSmalltau}a shows an example 
where the behaviors of
Eqs.~\eqref{eq_largetau1},\eqref{eq_intermediate1},\eqref{eq_largetau2} can be
seen in exact numerical calculations.  The small oscillations at large $\tau$
on top of Eqs.~\eqref{eq_largetau1},\eqref{eq_largetau2} are interference
effects, discussed later.

\section{Fast quenches}  \label{sec_smalltau}

We now consider small $\tau$, \emph{i.e.}, almost instantaneous quenches.  The
main observation is that the residual energy has the dependence
\begin{equation}  \label{eq_smalltau1}
\Delta{E} ~=~ \Delta{E}_{\rm inst.} ~-~ \eta \tau^2 ~+~ \ord(\tau^{\gamma})
\; ; \quad (\gamma>2) \, .
\end{equation}
%
%
The quadratic behavior is very robust, and is present for all values of $N$,
$\delta_0$, $K$ we have checked (Fig.~\ref{fig_AlltauSmalltau}a,b).  In
Fig.~\ref{fig_AlltauSmalltau}a, we can see the quadratic behavior (straight
line in log-log plot) for a $K{\ll}NU{\ll}\delta_0$ case, which is the
parameter region we analyze.  Fig.~\ref{fig_AlltauSmalltau}b shows quadratic
behaviors for cases outside this parameter regime.  

To explain this behavior, we start with $N=2$ bosons.  Writing the
wavefunction as $\ket{\psi(t)} = c_1(t)\ket{2,0}+ c_2(t)\ket{1,1}+
c_3(t)\ket{0,2}$, one can consider equations of motion for $c_i$, \emph{e.g.},
$\dot{c_1}(t) = -i \lbc{U}-\delta(t)\rbc c_1(t) + i\tfrac{K}{\sqrt{2}}
c_2(t)$, and solve for small orders in time:
\[
c_i(t) ~=~ c_i(0) ~+~ \dot{c_i}(0)t ~+~ \hf\ddot{c_i}(0)t^2 ~+~ O(t^3)  \; .
\]
Since $\dot{c_i}(0)$ is entirely imaginary, the linear terms in $t$ are
imaginary.  Linear-$t$ terms will thus cancel out from observable quantities
like $|c_i(t)|^2$ and $c_1(t)^*c_1(t)+c_1(t)c_1(t)^*$, which appear in the
expression for the final energy: 
$U\lba|c_1|^2+|c_3|^2\rba -
\frac{K}{\sqrt{2}}\lbc{c_1^*c_2+c_2^*c_3 + {\rm h.c.}}\rbc$. 
The energy at $\tau$ thus has a constant and a $\tau^2$ term, but no
$\ord(\tau)$ term.

For $N=2$, we can express the coefficient $\eta$ in Eq.~\eqref{eq_smalltau1}
analytically in terms of the initial $c_i(t=0)$, and also write analytic
expressions for $c_i(0)$, as solutions of cubic polynomials.  Unfortunately,
these expressions are too cumbersome to be useful.  For $\delta_0\gg{U},K$,
one can calculate perturbatively, yielding $\eta \sim K^2\delta_0/24$.

For $N>2$, the same argument holds for a leading $\tau^2$ correction.  For
very large $\delta_0$ and small enough $\tau$, one can use the approximation
that only the most imbalanced configurations are excited, and restrict to the
subspace $\ket{N,0}$, $\ket{N-1,1}$, $\ket{N-2,2}$.  The calculation is then
similar to the $N=2$ case.  One obtains 
\begin{equation}  \label{eq_smalltau2}
\eta \sim \frac{NK^2\delta_0}{48} \, , \qquad \Delta{E} ~\sim~ \Delta{E}_{\rm
  inst.}  - \frac{NK^2\delta_0}{48} \tau^2 \, ,
\end{equation}
at leading order in $\delta_0^{-1}$.  This also contains the $N=2$ result.
Fig.~\ref{fig_AlltauSmalltau}a,c show that this  expression works well for
$\delta_0\gg{NU}\gg{K}$.

%
While this compact result is valid only for $\delta_0{\gg}NU{\gg}K$, the level
crossing structure plays no role in this analysis, so in principle it can be
applied also to other parameter regimes; however the expressions for $\eta$
are too complicated to be useful in such cases.

%
Fig.~\ref{fig_AlltauSmalltau}a also displays the entire
instantaneous-to-adiabatic crossover, which includes our slow-quench and
fast-quench results but also a range of $\tau$ ($\sim$0.02 to $\sim$20) for
which we do not have simple descriptions.

\begin{figure}
\centering
 \includegraphics*[width=0.95\columnwidth]{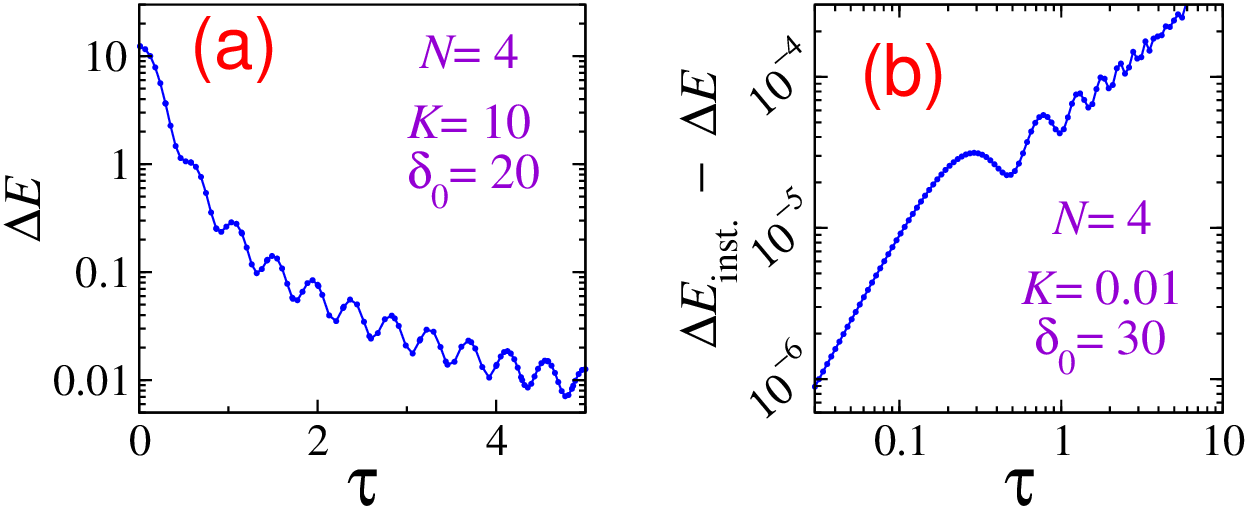}
\caption{ \label{fig_oscil}
(Color online.)  (a) Quantum interference at large $K$.
(b) Interferences resulting from using  $\ket{N,0}$ as initial state.
}
\end{figure}

\section{Quantum interferences}

Our treatment of multi-crossing configurations
(Fig.~\ref{fig_crossings_excessenergy}) utilized unit excitation probabilities
at the higher crossings, which is valid at small $K$.  However, if there is
substantial splitting at each crossing, the configurations offer rich
possibilities for quantum interference of different paths.  We have already
seen interference signatures in the mild oscillatory behavior in
Fig.~\ref{fig_AlltauSmalltau}a for ($N$,$K$) = (10,0.1) in the near-adiabatic
regime.  In Fig.~\ref{fig_oscil}a we show more pronounced
interference effects at large $K$.

Interference between paths also becomes prominent when the initial state is
not the ground state but has weights in more than one eigenstate, \emph{e.g.},
if one starts with the state $\ket{N,0}$  (Fig.~\ref{fig_oscil}b).
This is not exactly the ground state at $\delta=-\delta_0$ as long as
$\delta_0$ is finite.  It could however easily be an experimentally relevant
initial state.

\begin{figure}
\centering
 \includegraphics*[width=0.95\columnwidth]{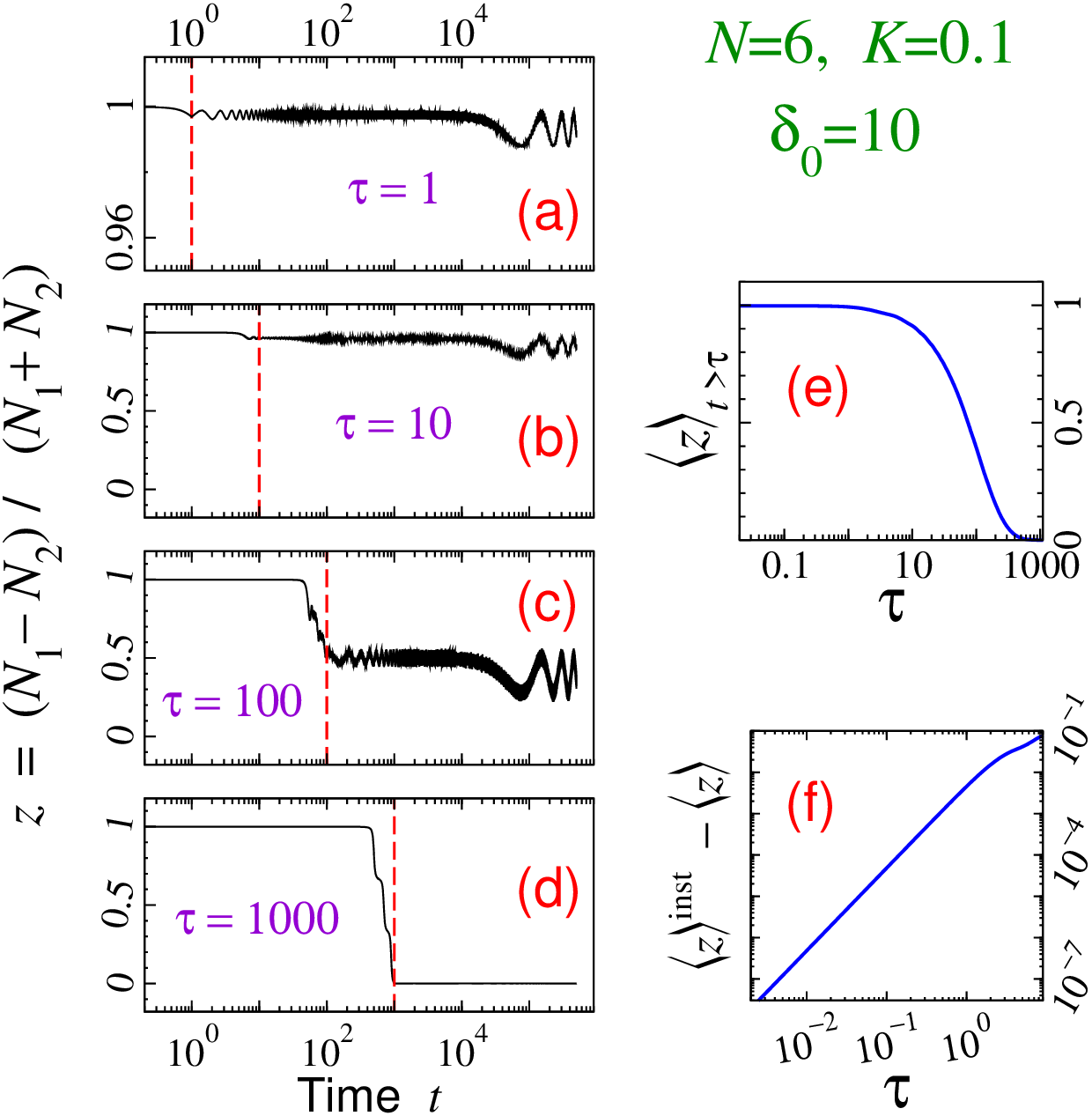}
\caption{ \label{fig_selftrapping}
(Color online.) 
Fate of self-trapping.  (a-d) Number imbalance dynamics during and after
quench, for various quench times.  Note different scale on top panel a.   
(e) Long-time average of relative number imbalance
  after quench is over at $t=\tau$.  Averaging is performed between $t=\tau$
  and $t=t_{\rm up} = 500000$. 
(f) Quadratic deviation of $\xpct{z}_{t>\tau}$ from instantaneous case, for
small $\tau$.
}
\end{figure}

\section{Self-trapping with finite-rate quenches}  \label{sec_selftrapping}

We now consider the effect of finite quench times on the quantum self-trapping
phenomenon. \cite{MilburnCorneyWrightWalls_PRA97,
RaghavanSmerziKenkre_PRA99,RaghavanSmerziFantoniShenoy_PRA99,
KalosakasBishopKenkre_JPhysB03,KalosakasBishopKenkre_PRA03,
Salgueiro-etal_EPJD07, TonelLinksFoerster_JPA05} 
Self-trapping involves dynamics starting with a biased state, which is the
case for an instantanous $\tau=0$ quench from large negative
$\delta=-\delta_0$.  This is the type of dynamics analyzed in Ref.\
\onlinecite{Salgueiro-etal_EPJD07}.  The relative number imbalance
$z=\xpct{n_1-n_2}/N$ can oscillate around a nonzero value of $z$ if $NU/K$ and
the starting $z$ are large enough.
Since the state $\psi_{t=\tau}$ after a finite-$\tau$ quench is closer to the
unbiased ($z=0$) ground state at $\delta=0$, self-trapping is clearly
weakened at larger $\tau$.

We can characterize self-trapping through the long-time average of
$z=\xpct{n_1-n_2}/N$ at times $t>\tau$.  A nonzero $\xpct{z}_{t>\tau}$
indicates self-trapping.  The definition is somewhat subtle because quantum
self-trapping involves nonzero $\xpct{z}_{t>\tau}$ up to some very large time
but not really infinite times.  (This very large timescale is due to the tiny
splitting energies of the higher level crossings in Fig.\
\ref{fig_crossings_excessenergy}a. \cite{MilburnCorneyWrightWalls_PRA97,
RaghavanSmerziKenkre_PRA99,RaghavanSmerziFantoniShenoy_PRA99,
KalosakasBishopKenkre_JPhysB03,KalosakasBishopKenkre_PRA03,
Salgueiro-etal_EPJD07, TonelLinksFoerster_JPA05})
The long-time average therefore depends on the upper time limit $t_{\rm up}$
up to which $z(t)$ is averaged, and will always be zero for large enough
$t_{\rm up}$.  This is in contrast to the classical self-trapping observed in
the two-site discrete nonlinear Schr\"odinger equation, where the
`self-trapped' states are truly stationary states.  \cite{ClassicalSelfTrapping}

In Fig.~\ref{fig_selftrapping}a-d we demonstrate the weakening of the
self-trapping effect at larger $\tau$, through the behavior of $z(t)$.  The
logarithmic scale highlights the fact that there are several imporant time
scales in the dynamics.  The intermediate-$\tau$ cases have more complicated
wavefunctions at the end of the quench, as analyzed in Section
\ref{sec_largetau}.  Hence the dynamics is most interesting for intermediate
$\tau$.  This can be seen for example through a larger number of peaks in the
Fourier transform of the intermediate-$\tau$ time evolution data.  In
comparison, the $\tau\lesssim{1}$ and $\tau\gg{100}$ cases show fewer
high-frequency components in the dynamics.


Fig.\ \ref{fig_selftrapping}e shows the dependence of $\xpct{z}_{\tau<t<t_{\rm
up}}$ on $\tau$.  The averaging displayed in Fig.\ \ref{fig_selftrapping}e,f
was performed up to $t_{\rm up} =5\times10^5$; the $z(t)$ behaviors in
Fig.~\ref{fig_selftrapping}a-d suggests this is a reasonable value for
capturing the self-trapping phenomenon.  The behavior of $\xpct{z}$ is similar
to the $\tau$-dependence of the residual energy $\Delta{E}$
(Fig.~\ref{fig_energyEvolution}e); howerver the ambiguity in the averaging
procedure makes it difficult to perform a systematic analysis similar to what
we have presented for $\Delta{E}$ in previous sections.  The similarity of
Fig.~\ref{fig_energyEvolution}e and Fig.~\ref{fig_selftrapping}e is not
surprising because both the $z(t>\tau)$ dynamics and the final energy are
determined by the state at the end of the quench, \emph{i.e.}, the weight of
excited states in $\psi_{t=\tau}$.  At small $\tau$ there is also a quadratic
deviation of $\xpct{z}_{t>\tau}$ from the instantaneous case
(Fig.~\ref{fig_selftrapping}f), as there is for $\Delta{E}$.

\section{Summary \& Open issues} 

This work analyzes finite-rate quenches most relevant to the self-trapping
phenomenon, namely, from $\delta=-\delta_0$ to $\delta=0$ with parameters
$K{\ll}NU{\ll}\delta_0$.
While quenches of the tilt $\delta$ are natural for the dimer, it is
relatively new in the non-equilibrium literature because bias quenches do not
appear naturally in the many-site case.
We provide concise analytical results for the residual energy in
parametrically different regimes of small, intermediate, and large $\tau$
(Eqs.\ \ref{eq_smalltau2}, \ref{eq_intermediate1}, \ref{eq_largetau2}).  The
small $\tau$ result \eqref{eq_smalltau2} is perturbative in $\tau$, while
Eqs.\ \ref{eq_intermediate1}, \ref{eq_largetau2} are obtained from the
Landau-Zener formula and are thus non-perturbative, despite the linear
dependence \eqref{eq_intermediate1} at intermediate $\tau$.

The importance of Bose-Hubbard dimer dynamics reaches far beyond the cold-atom
context, as the model appears in diverse areas of physics.  For example, it is
equivalent to a large-spin Hamiltonian:
$H = -KJ_x + {U}J_z^2 + {\delta}J_z$.
This describes single-molecule magnets.  However for molecular magnet
experiments the anisotropy ($J_z^2$) term usually has negative coefficient,
and effects of host lattice vibrations or nuclear spins cannot always be
neglected.

The residual energy is in principle experimentally measurable in cold-atom
realizations through time-of-flight measurements that provide energy
information from the cloud expansion rate (\emph{e.g.}, Ref.\
\onlinecite{Rempe_PRA04}).
For molecular magnet experiments energy is difficult to measure, but
magnetization dynamics (analogous to our $z(t)$ dynamics) is commonly
reported.

Our work raises a number of open issues.  There are a number of parameter
regimes other than ours which might be of interest.  Examples are other values
of initial and final $\delta$, or of $K/U$.  Refs.\
\onlinecite{WitthautGraefeKorsch_PRA06,
Smith-MannschottChuchemHillerKottosCohen_PRL09} have considered sweeps of
$\delta$ from negative to positive infinity, for parameter regions and initial
states quite different from ours.  Clearly, we are only seeing the beginning
stages of an emerging unified dynamical picture.  Another intriguing issue is
the connection to the mean-field description via the discrete Gross-Pitaevskii
equation.  Ref.\ \onlinecite{WitthautGraefeKorsch_PRA06} has worked out some
Landau-Zener issues for the mean-field dimer, for $U<0$ and
$(-\infty\rightarrow+\infty)$ sweeps.  Numerical explorations for
$(-\delta_0\rightarrow0)$ quenches have shown us behaviors similar to what we
have presented for the full quantum case.  Finally, the present results may
need to be adapted for specific realizations,
once such experiments are designed.

%
%


\end{document}